# Aluminum Nitride Two-Dimensional-Resonant-Rods

Xuanyi Zhao, Luca Colombo, and Cristian Cassella

Northeastern University, Boston (MA) - USA

*In the last decades, Bulk-Acoustic-Wave (BAW) filters have been essential components of 1G-to-4G radios. These devices rely on the high electromechanical coupling coefficient ($k_t^2 \sim 7\%$), attained by Aluminum Nitride (AlN) Film-Bulk-Acoustic-Resonators (FBARs), to achieve a wideband and low-loss frequency response. As the resonance frequency of FBARs is set by their thickness, the integration of multiple FBARs, to form filters, can only be attained through the adoption of frequency tuning fabrication steps, such as mass loading or trimming. However, as the ability to reliably control these steps significantly decays for thinner FBARs, manufacturing FBARs-based filters, addressing the needs of emerging IoT and 5G high-frequency applications, is becoming more and more challenging. Consequently, there is a quest for new acoustic resonant components, simultaneously exhibiting high-$k_t^2$ and a lithographic frequency tunability. In this work, a novel class of AlN resonators is presented. These radio-frequency devices, labelled as Two-Dimensional-Resonant-Rods (2DRRs), exploit, for the first time, the unconventional acoustic behavior exhibited by a forest of locally resonant rods, built in the body of a profiled AlN layer that is sandwiched between a bottom un-patterned metal plate and a top metallic grating. 2DRRs exhibit unexplored modal features that make them able to achieve high-$k_t^2$, a significant lithographic frequency tunability and a relaxed lithographic resolution, while relying on an optimal AlN crystalline orientation. The operation of 2DRRs is discussed, in this work, by means of analytical and finite-element (FE) investigations. The measured performance of the first fabricated 2DRR, showing a $k_t^2$ in excess of 7.4%, are also reported.*

In the last decades, acoustic resonators[1–4] and filters have represented key components for several radio-frequency (RF) applications and systems. For instance, their superior performance, when compared to conventional electromagnetic counterparts, have made them essential to form frequency selective passive components in miniaturized RF platforms. Aluminum Nitride (AlN) Film-Bulk-Acoustic-Resonators (*FBARs*)[2,5,6] have been extensively used to form filters in commercial RF front-ends. In fact, thanks to their ability to attain large electromechanical coupling coefficient ($k_t^2 \sim 7\%$), while being manufacturable through conventional semiconductor fabrication processes, FBARs have enabled commercial ultra- and super-high-frequency (U/SHF) acoustic filters, exhibiting fractional bandwidths and performance exceeding those required by *4G* communication systems. However, despite their high-$k_t^2$, the resonance frequency ($f_{res}$) of FBARs is set by the thickness of the different forming layers[7,8]. This feature renders the manufacture of FBARs, having different $f_{res}$-values, only attainable through post-processing steps, like *trimming* and *mass loading*, thus leading to a much higher fabrication complexity. Such complexity becomes even more significant when FBAR-based filters, operating at higher frequencies, are required to satisfy the needs of the emerging super- and extremely-high-frequency (*S/EHF*) *IoT* and *5G* applications[9–12]. In fact, in such case, FBARs with significantly thinner metallic and piezoelectric layers are required, to enable the desired higher frequency operation. This critical constraint comes with an increased sensitivity of $f_{res}$ with respect to the thickness of the FBARs layers. This feature renders any step of mass-loading or trimming not easily controllable and, consequently, hardly usable in a large-scale production. Therefore, in recent years, many groups have researched novel AlN-based device technologies, simultaneously enabling high-$k_t^2$ and a lithographic frequency tunability[13–21]. In particular, Cross-Sectional-Lamé-Mode resonators (CLMRs)[13,22] and Two-Dimensional-Mode-Resonators (2DMRs)[16] were recently demonstrated. CLMRs excite a combination of vertical ($S_1$) and lateral ($S_0$) longitudinal motions[23], in AlN plates, through a coherent combination of the $d_{31}$ and $d_{33}$ piezoelectric coefficients. In contrast, 2DMRs excite a set of dispersive $S_1$-Lamb wave modes, confined between the strips forming their metallic gratings. While CLMRs enable a comparable $k_t^2$-value attained by FBARs and a significant lithographic frequency tunability ($\Delta f$), conventional 2DMRs can generally achieve a slightly lower $k_t^2$ (<5%) and a reduced $\Delta f$-value. However, these devices can excite resonant vibrations through metallic gratings that are formed by wider metallic strips than those required by CLMRs, operating at the same frequency. For this reason, they enable significantly lower ohmic losses than CLMRs, hence higher *Q*, thus being promising candidates to achieve monolithic integrated acoustic filters, for *5G* communication systems. Only recently, modified 2DMRs[17], using a set of top and bottom metallic frames, were proposed to enable comparable $k_t^2$ and $\Delta f$-values attained by CLMRs, while still ensuring a more relaxed lithographic resolution. However, the adoption of these frames leads to a heavily enhanced fabrication complexity, with respect to conventional 2DMRs and CLMRs. Interestingly, few years ago, *Zuo, Southin* et al.[24,25] investigated, through numerical methods, the operational features relative to a set of rod-modes, labeled here as dilatational modes. These modes can exhibit superior $k_t^2$ (~10%, in AlN) and $\Delta f$-values. However, up to date, it has been believed that such a high-$k_t^2$ is only attainable in one isolated and narrow rod. Consequently, the use of dilatational modes has been considered not suitable for practical filtering applications, where devices with large input capacitances ($C_0$) are required to ensure proper functioning with *50Ω*-matched electronics. In fact, there has been no proposition of any multi-finger acoustic resonator design, capable to efficiently and coherently excite such modes, in single piezoelectric slabs.

In this work, a new class of multi-finger acoustic resonators is presented. These devices are labeled as *Two-Dimensional-Resonant-Rods* (*2DRRs*). 2DRRs are formed by a profiled AlN-layer, sandwiched between one top metal grating and a grounded bottom metal plate (Figure 1-a). They exploit, for the first time, the excitation of a combined set of dilatational modes, in a single AlN plate. To do so, 2DRRs rely on a *forest* of locally resonant piezoelectric rods, built on the top surface of a thin multi-layered beam. These rods are attained by partially etching the AlN-portions included between adjacent metal strips forming the grating. The etching profile is engineered in order to form steep trenches that exhibit an evanescent lateral wavevector component ($k_x$), at $f_{res}$. This feature permits to confine the resonant vibration within the rod, thus rendering adjacent rods only weakly and reactively coupled[26], so as to ensure a common frequency of operation. Moreover, because of their modal characteristics, 2DRRs exhibit a higher sensitivity of $f_{res}$ with respect to the rod-width (*a*). Ultimately, the adoption of trenches allows to suppress any non-vertical electric field line that would otherwise be generated between adjacent strips forming the grating, thus lowering the obtainable $k_t^2$. Meanwhile, by adopting un-patterned bottom metal plates instead of patterned ones, like those required by both CLMRs and 2DMRs to achieve high-$k_t^2$, 2DRRs can rely on AlN-films exhibiting an optimal crystalline orientation, even when thinner films are needed to operate in the *5G* spectrum[27]. Here, we report the performance of the first fabricated 2.3GHz 2DRR (Figure 1) along with discussions about its operation, through both a simplified one-dimensional analytical model and through finite-element-methods (FEM).

To explain how multiple dilatational modes can be excited in the reported 2DRR, we can study the acoustic propagation characteristics exhibited by one of its periodic cells (*i.e.* the *unit-cell*, see Figure 1). The unit-cell is formed by two main regions, here defined as *trench* and *rod*. The trench is formed by a bottom metallic plate and by a thin AlN layer. The rod is formed by a thicker AlN-film and by a second metallic layer. In the following, the thicknesses of the thin AlN-layer, of the thicker AlN-layer, of the bottom metal layer and of the top-metal layer will be labeled as $T_{AlN}^{(1)}$, $T_{AlN}^{(2)}$, $T_m^{(1)}$ and $T_m^{(2)}$, respectively. Also, the length of the unit-cell will be labeled as *L*. Moreover, we will refer to the interface between the rod and the trench and to the trench portion underneath each rod as *S* and *Region A*, respectively. It is easy to realize that the top face of each rod behaves as a stress-free (*SF*) boundary. Because of the distributed nature of the rod, this boundary translates into a different mechanical boundary condition *(B.C)* across *S*, at different frequencies. Such *B.C* can significantly perturb the vertical displacement in the rod, $u_z(z)$, generated by any force ($F(x, z = 0)$) applied, to the rod, from Region A through *S*. In particular, when assuming that only a negligible dispersion affects thickness-extensional (*TE*) waves in the rod, $F(x, z = 0)$ can be considered uniform across *S*, thus being simply indicated as $F$. The determination of $F$ (Eq. (*S5*)) is particularly important as it allows to compute the value of the driving impedance ($Z_b$) relative to the rod (see Eqs. (*S1-S5*)). $Z_b$ allows to establish the influence of the rod on the *B.C* exerted by Region A, across S. The distribution of $Z_b$ *vs. f*, for our fabricated 2DRR, is plotted in the Supplementary Material (Figure S2). As the trench ($T_{AlN}^{(1)}+T_m^{(1)}$) is significantly thinner than the rod ($T_{AlN}^{(2)}+T_m^{(2)}$), it is reasonable to assume, in a first order of approximation, that only the flexural ($A_0$) and the lateral ($S_0$) plate modes can propagate within the trench[23]. However, as the coupling between the rod and the trench can only occur through vertical fields and since a low dispersion affects the velocity of the $S_0$-mode, for the thickness over lambda ($\lambda$) ratio used for the trench[21], we neglect any coupling, through the $S_0$, between the rod and the trench. This simplification allows to consider the $A_0$ as the only existing propagating mode that can guide acoustic energy between adjacent unit-cells of 2DRRs and permits to assume a uniform transversal displacement ($v(x)$), throughout the thickness of the trench. In this scenario, $v(x)$ can be estimated by solving a one-dimensional (*1D*) Euler-Bernoulli equation of motion (Eq. (1)), after selecting a proper set of *B.Cs*[22].

$$E_t I_t \frac{d^4 v(x)}{dx^4} - \rho_t A_t \omega^2 v(x) = -F\left(H\left(x - \frac{a}{2}\right) - H\left(x + \frac{a}{2}\right)\right) \quad (1)$$

In Eq. (1), $E_t$, $I_t$ and $A_t$ are, respectively, the effective Young's modulus (Eq. S7), second moment of inertia and cross-sectional area relative to the trench. The function $H$ is the Heaviside function. From the homogenous of Eq. (1), it is straightforward to estimate the real dispersive wavevector (*k*, see Eq. (*S6*)) associated to the $A_0$-mode, when excluding the presence of the rods. The distribution of $v(x)$ can be derived through the same methodology introduced in [22]. In particular, let *t*(x) and *w*(x) represent *v*(*x*), for the left and right sides of Region A (see Figure 1c). Both displacement distributions can be expressed as the superposition of left/right propagating and evanescent waves (Eqs. (2-3)). It is necessary to point out that the evanescent not propagating decaying terms can only be ignored from the solution of Eq. (1) when dealing with uniform plates, thus not including any rod.

$$t(x) = t_l(x)e^{-ikx} + t_r(x)e^{ikx} + t_{re}(x)e^{-kx} + t_{le}(x)e^{kx} \quad (2)$$

$$w(x) = w_l(x)e^{-ikx} + w_r(x)e^{ikx} + w_{re}(x)e^{-kx} + w_{le}(x)e^{kx} \quad (3)$$

In Eqs. (2-3), the subscript *l* and *r* indicate the moving directions of the different components (from left-to-right and from right-to-left, respectively). The subscript *e* refers to the evanescent wave components. Investigating the wave transmission through the unit-cell requires the computation of a transmission-matrix, $[T]_{4x4}$ (see Eq. (*S23*)), relative to the displacement field moving from one edge of the unit-cell towards the other. $[T]_{4x4}$ maps the relationship between the amplitudes of the different components forming Eqs. (2-3), after considering the transformation that such components undergo, when moving through the trench portions adjacent to Region A. From $[T]_{4x4}$, it is easy to determine the transmission coefficient, *T*, for the propagating displacement component of *v*(x), leaving one edge of the unit-cell towards an infinite number ($N \to \infty$) of cascaded identical unit-cells. In particular, *T* is expected to be unitary at frequencies at which the rod does not affect the propagating features of the unit-cell. In contrast, *T* is expected to approach zero at the frequencies ($f^{(n)}$) at which the rod exhibits the largest influence. This important feature is determined by a process of acoustic energy storage in the rods and in Regions A. This reactive phenomenon prevents the flow of real power from adjacent unit-cells. The extrapolation of *T* allows to identify the existence of *passbands* and *stopbands* for the propagation of the $A_0$-mode in 2DRRs. It is worth emphasizing that the adoption of an infinite sequence of periodic cells, during the evaluation of *T*, permits to neglect the *edge* effect that, for a finite *N*-value, can partially alter the validity of our analytical treatment. The expression of *T* is rather cumbersome and its frequency distribution can only be determined numerically. As an example, a widespan representation of *T* vs. *f*, relative to the fabricated 2DRR, is plotted in Figure *S3*. As evident, multiple stopbands exist for the $A_0$ propagation in the analyzed structure. In favor of a clearer visualization, the same *T* distribution, along with the corresponding attenuation coefficient (*R*=1-|*T*|), is plotted in Figure 2-b, for close frequencies

to the experimentally measured $f_{res}$ (~2.35GHz). In order to fully understand the origin of the stopbands, we can look at both the phase ($\varphi$) and the real part ($\Gamma_{real}$) of the reflection coefficient ($\Gamma$) relative to the propagating displacement components, at the right edge of the unit-cell. The distributions of $\varphi$ and $\Gamma_{real}$ *vs. f* are reported in Figure 2. As evident, $\Gamma$ exhibits a sequence of resonance conditions corresponding to the *f*-values at which $\varphi$ is equal to $\pm\pi$. Some of them (the *series resonances*) correspond to $\Gamma_{real}$-values equal to -1 (*i.e.* the expected value for stress-free boundaries) whereas the remaining ones (the *parallel resonances*) correspond to $\Gamma_{real}$-values equal to 1 (*i.e.* the expected value for fixed boundaries). These latter resonances identify the frequencies at which the rod is expected to exhibit the largest influence on the propagation characteristics of the unit-cell. The existence of multiple un-correlated frequencies at which such strong interaction exists is caused by the dispersive characteristics of the $A_0$-mode. Therefore, within the stopbands, the $A_0$-mode exhibits a large and evanescent wavevector ($k_{ef}$ =N·$k_{ef}^{(i)}$, being $k_{ef}^{(i)}$ the wavevector relative to one arbitrary unit-cell) that prevents the exchange of acoustic energy between adjacent unit-cells. In order to demonstrate the evanescent behavior of $k_{ef}$, we report, in Figure 2-c, the numerically found real ($k_{ef\text{-real}}$) and imaginary ($k_{ef\text{-im}}$) parts of $k_{ef}^{(i)}$, for the analyzed and built 2DRR. Evidently, within the stopband, $k_{ef}^{(i)}$ is purely imaginary, which is a direct proof that no propagation of the $A_0$ occurs within this frequency range.

Despite the fact that one side of an arbitrary unit-cell was used as the reference location ($x_0$) for the computation of *T*, the magnitude of *T* is invariant to $x_0$. So, the same *T*-values would be attained if a different reference location, included in the Region A, were used. This important consideration is crucial to understand the origin of the unique modal features that characterize the operation of 2DRRs (see the FEM simulated displacement modeshape in Figure 3-a). In fact, the resonant vibration of these devices is piezoelectrically generated, in the rods and in Regions A, from the vertical electric field ($E_z$) that exists between the top metal strip and the bottom metal plate. In particular, $E_z$ couples to mechanical strain through the AlN $d_{31}$ and $d_{33}$ piezoelectric coefficients. However, because of the described dispersive properties of the unit-cell, the lateral edges of Regions A, from which the acoustic energy would tend to leak towards adjacent unit-cells, behave as fixed-boundaries. Consequently, the generated acoustic energy comes to be stored in the rod structures, whose lateral sides act as *SF* boundaries, hence being more prone to deform. As experimentally demonstrated (see Sec. SII), this unique operational feature allows to generate more mechanical energy than possible when no trench is used, thus being ultimately the main responsible for the high-$k_t^2$ attained by 2DRRs.

The electrical performance of the fabricated 2DRR (details on the fabrication flow in Sec. SIII) were extracted (Figure ) through conventional RF characterization tools. This device, which is formed by 20 unit-cells, shows measured $k_t^2$, resistance at resonance ($R_{tot}=R_m+R_s$), loaded quality factor $Q_{3dB}$ (extracted from the 3dB bandwidth) and $C_0$ in excess of 7.4%, 56Ω, 185 and 325fF (corresponding to an impedance of *208Ω*), respectively. The measured $k_t^2$ and $C_0$ values match closely their FEM predicted values (7.7% and 300fF, respectively). Ultimately, the capability to lithographically define the resonance frequency of 2DRRs was also investigated through FEM (Figure 3-c). This was done by simulating the trends of $k_t^2$ and $f_{res}$ vs. $a$, when considering the same material stack adopted for the reported 2DRR device. As evident, 2DRRs simultaneously enable a significant lithographic frequency tunability ($\Delta f$>117MHz) and a large $k_t^2$ exceeding 5%. This unique feature renders them promising components to form monolithic integrated wideband filters, for next-generation RF front-ends.

**Figures and Figure Captions**

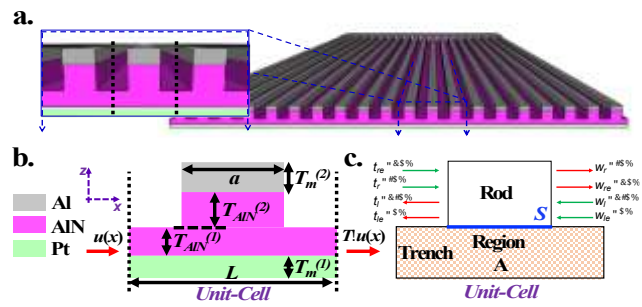

Figure 1: a) Schematic view of the 2.3GHz 2DRR reported in this work; b) A cross-sectional view of the periodic unit-cell forming the device and description of the materials adopted in this first 2DRR implementation ($T_{AlN}^{(1)}$=400nm, $T_{AlN}^{(2)}$=600nm, $T_m^{(1)}$=250nm and $T_m^{(2)}$=330nm); c) Adopted nomenclature for the three main regions forming the 2DRR unit-cell and schematic representation of the different displacement components defined in Eqs. (2-3).

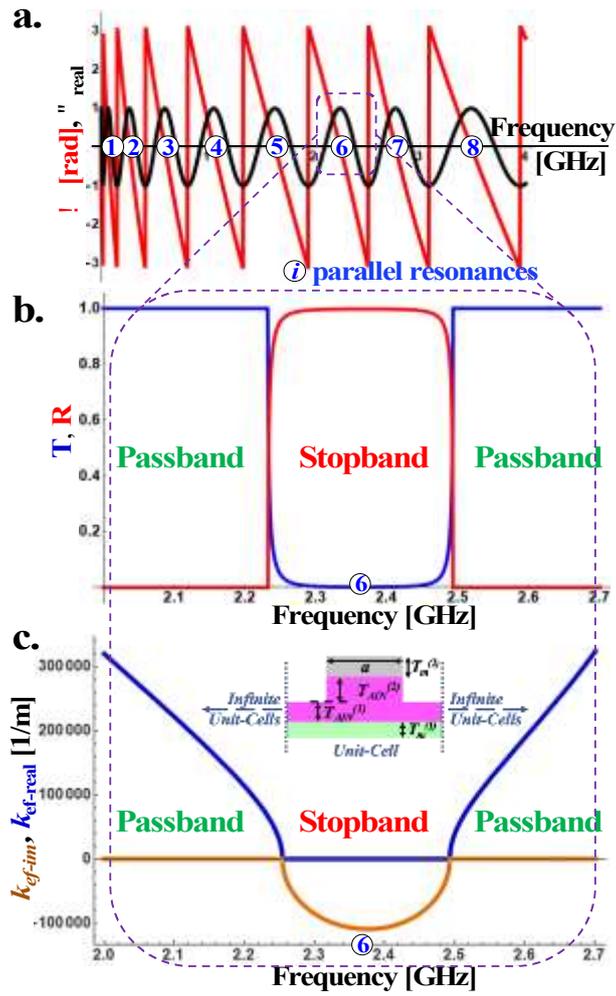

Figure 2: a) Simulated trends of $\varphi$ and $\Gamma_{real}$ vs. $f$ and relative to the unit-cell of our fabricated 2DRR; b) analytically derived trends of $T$ (in blue) and $R$ (i.e. 1-|T|, in red) vs. $f$, relative to the fabricated 2DRR (Figure 1) and for $f$ varying around the measured $f_{res}$; c) Analytically derived $k_{ef\text{-real}}$ (in blue) and $k_{ef\text{-im}}$ (in brown), when assuming the same geometrical and material characteristics adopted in the fabricated device.

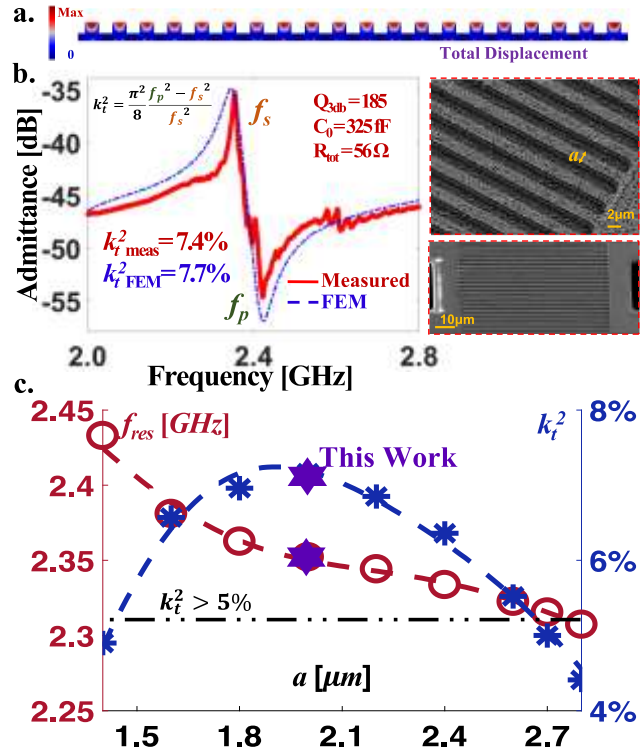

Figure 3: a) FEM simulated resonant total displacement modal distribution, relative to our fabricated 2DRR. b) Measured (in red) and simulated (in blue) admittance for the fabricated device. As evident, with the exception of few addition spurious resonances in the measured results, there is a close matching between measured and FEM simulated results. Scanned Electron Microscope (SEM) pictures of the fabricated device are also shown. c) FEM simulated distributions of $f_{res}$ and $k_t^2$ vs. $a$, when assuming the same material stack and geometrical parameters used for the fabricated 2DRR device (Figure 1).

The data that support the findings of this study are available from the corresponding author upon reasonable request.

**Aluminum Nitride Two-Dimensional-Resonant-Rods (*Supplementary Material*)**

*Xuanyi Zhao, Luca Colombo, and Cristian Cassella*

**SI. Analytical Study of the 2DRR Unit-Cell**

In order to analyze the operation and unconventional dispersive characteristics of 2DRRs, it suffices to investigate the acoustic behavior relative to one of their periodic cell (*i.e.* the *unit-cell*, see Figure S1). In fact, such cell captures all transitions between different acoustic characteristics that periodically occur across the entire device geometry. As discussed in the main manuscript, the unit-cell is formed by two main regions, defined as *trench* and *rod*, which are characterized, for the 2DRR fabricated in this work, by the mechanical and geometrical parameters shown in Figure S1.

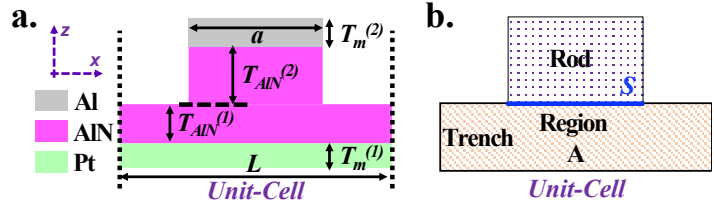

Figure S1:a) cross-sectional schematic relative to each unit-cell forming the 2DRR experimentally demonstrated in this work; b) schematic representation and relative nomenclature relative to the main parts forming the unit-cell of the 2DRR reported in *a*.

The faces of each rod that are orthogonal to their main vibrational direction (vertical *z*-direction) are characterized by different boundary conditions (*B.Cs*). The top face behaves as a stress-free boundary, whereas the bottom face is directly attached to the trench region labeled as *Region A* (see Figure S1). In particular, this latter face is loaded by the longitudinal force (*F* see Eq. (1) in the main manuscript) originated in Region A, oriented along the *z*-direction and perturbing the stress distribution and propagation features relative to the entire unit-cell. When assuming that only a negligible dispersion affects the velocity of longitudinal thickness-extensional (*TE*) waves in the rod, $u_z(z)$ can be estimated, in its frequency-domain representation, as (Eq. (S1))[1]:

$$u_z(z) = -\frac{F}{\omega \rho_{rod} A_{rod} c_{rod}} \left( \sin(k_{rod} z) + \frac{\cos(k_{rod} z)}{\tan\left(k_{rod}\left(T_{AlN}^{(2)} + T_m^{(2)}\right)\right)} \right) \quad (S1)$$

where *F* can be rewritten as $-E_{rod} A_{rod} \varepsilon_z(z=0)$, being $\varepsilon_z(z=0)$ the strain across *S*. In Eq. (1) $\rho_{rod}$, $A_{rod}$ and $c_{rod}$ represent the effective mass density, the cross-sectional area (*i.e.* $(T_{AlN}^{(2)} + T_m^{(2)}) \cdot W$, being *W* the out-of-plane dimension relative to both the trenches and the rods) and the nondispersive phase velocity, for longitudinal waves, in the rods. In contrast, $k_{rod}$ and $\omega$ represent the wavevector relative to the same vertical motion and the natural frequency (*i.e.* $\omega = 2\pi f$, being *f* the frequency), respectively. In order to estimate the mechanical properties and resonance frequency ($f_{res}$) of the *TE*-mode, in the rod, $\rho_{rod}$ and $c_{rod}$ can be found after computing effective values for the Young's modulus ($E_{rod}$) and mass density, based on the geometrical and mechanical parameters relative to the materials forming the rods. Therefore, $E_{rod}$ and $\rho_{rod}$ can be estimated as (Eqs. (2-3)):

$$E_{rod} = \frac{\left(E_{AlN} T_{AlN}^{(2)} + E_m^{(2)} T_m^{(2)}\right)}{T_m^{(2)} + T_{AlN}^{(2)}} \quad (S2)$$

$$\rho_{rod} = \frac{\left(\rho_{AlN} T_{AlN}^{(2)} + \rho_m^{(2)} T_m^{(2)}\right)}{T_m^{(2)} + T_{AlN}^{(2)}} \tag{S3}$$

From Eqs. (2-3), $c_{rod}$ can be found as:

$$c_{rod} = \sqrt{\frac{E_{rod}}{\rho_{rod}}} \tag{S4}$$

The driving impedance across $S$ ($Z_b$), relative to each rod, can be found through the Mason formalism[2] (Eq. (S5)).

$$Z_b = \frac{F}{vel(z=0)} = \frac{F}{-i\omega u_z(z=0)} = -i\, \rho_{rod} A_{rod} c_{rod} \tan\left(k_{rod}\left(T_{AlN}^{(2)} + T_m^{(2)}\right)\right) \tag{S5}$$

In Eq. (S5), $vel(z=0)$ is the magnitude of the laterally uniform vertical velocity (*i.e.* time derivative of $u_z(z=0)$ with respect to time), at $S$. It is straightforward to notice (Figure S2) that $Z_b$ exhibits both a local maximum and a local minimum, at two correlated frequencies, $f_{min}$ and $f_{max}$, respectively. In particular, for $f$ equal to $f_{min}$, $Z_b$ is equal to zero. Thus, at this frequency of operation, the rod does not exert any constraint on the displacement at $S$. For this reason, $S$ acts, at $f_{min}$, as a conventional *SF* boundary, placed in the active resonator portion. In contrast, for $f$ equal to $f_{max}$, the rod imposes a virtual *fixed*-constraint across $S$. It is worth mentioning that other non-conventional *B.C*s characterize the impact of the rod on the behavior of $S$ and, consequently, of Region A, for different frequencies from $f_{min}$ and $f_{max}$. We report (Figure S2) the distribution of $Z_b$, for the material and geometrical characteristics reported in Figure 1 of the main manuscript.

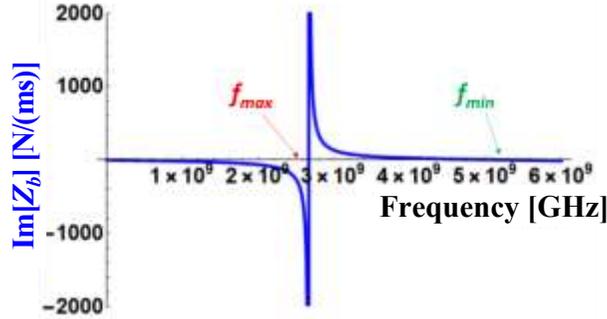

Figure S2: Analytically simulated distribution of the imaginary part of $Z_b$ as the frequency (*f*) is varied between 0 and 6GHz.

As discussed in the main manuscript, the propagation characteristics within the unit-cell can be found by solving the *1D* Euler-Bernoulli equation (see Eq. (1) in the main manuscript), in terms of the transversal displacement in the trench ($v(x)$), after properly selecting a suitable set of *B.C*s. However, from the solution of the homogenous of Eq. (1), it is useful to extract the wavevector (Eq. (S6)) associated to the $A_0$-mode, when neglecting the presence of the rods.

$$k = \frac{\sqrt{2}\, 3^{1/4} \sqrt{\omega}\, \rho_t^{1/4}}{\sqrt{T_m^{(1)} + T_{AlN}^{(1)}}\, E_t^{1/4}} \tag{S6}$$

In Eq. (S6), $E_t$ and $\rho_t$ are effective Young's modulus and mass density relative to the trench. In analogy to what done for the rods (Eqs. (S2-S3)), these parameters can be found, for the unit-cell of the fabricated 2DRR (Figure S1), as (Eqs. (S7-S8)):

$$E_t = \frac{\left(E_{AlN}T_{AlN}{}^{(1)} + E_{Pt}T_m{}^{(1)}\right)}{T_m{}^{(1)} + T_{AlN}{}^{(1)}} \tag{S7}$$

$$\rho_t = \frac{\left(\rho_{AlN}T_{AlN}{}^{(1)} + \rho_{Pt}T_m{}^{(1)}\right)}{T_m{}^{(1)} + T_{AlN}{}^{(1)}} \tag{S8}$$

It is important to point out that the magnitude of *F*, in Eq. (1) of the main manuscript, can be directly expressed in terms of the driving impedance of the rod ($Z_b$, see Eq. (*S5*)). In fact, its value can be computed as $F= Z_b \cdot vel$(z=0), thus being independent from *x*. The distribution of $v(x)$ can be derived through the methodology discussed in [1]. In particular, after breaking *v*(x) into its portions (*t*(x) and *w*(x)), for the left and right sides of Region A (see Eqs. (2-3) in the main manuscript), a scattering matrix ($[G]_{4x4}$) can be defined. $[G]_{4x4}$ captures the changes of the wave characteristics (Eq. (*S9*)) that occur at the transitions between each uncovered trench region and Region A (see Figure S1). More specifically, $[G]_{4x4}$ allows to map the interaction between the wave components going towards the rods ($t_l$, $t_{le}$, $w_r$, $w_{re}$) and those ($t_r$, $t_{re}$, $w_l$ and $w_{le}$) that, instead, are reflected by them (Eq. (*S9*)). For this reason, it is a function of the geometrical and material composition of the entire unit-cell. The matrix $[G]_{4x4}$ is reported in Eq. (*S10*).

$$\begin{bmatrix} t_l \\ t_{le} \\ w_r \\ w_{re} \end{bmatrix} = [G]_{4x4} \begin{bmatrix} t_r \\ t_{re} \\ w_l \\ w_{le} \end{bmatrix} \tag{S9}$$

$$[G]_{4x4} = \begin{bmatrix} r & r_{ef} & t & t_{ef} \\ r_{fe} & r_e & t_{fe} & t_e \\ t & t_{ef} & r & r_{ef} \\ t_{fe} & t_e & r_{fe} & r_e \end{bmatrix} \tag{S10}$$

It is important to point out that the two vectors shown in the left- ($[v_1]_{4x1}$) and right- ($[v_2]_{4x1}$) sides of Eq. (*S9*) are composed by the amplitudes of the different wave components forming *t*(x) and *w*(x), at the lateral edges of Region A. Regarding $[G]_{4x4}$, *r* and *t* map the reflection and transmission coefficients for the different propagating wave components that are incident towards Region A. Clearly, thanks to the symmetric nature of this problem, *r* and *t* have the same value for both *t*(x) and *w*(x). Similarly, $r_e$ and $t_e$ represent the reflection and transmission coefficients relative to the evanescent terms of *t*(x) and *w*(x). Ultimately, $r_{ef}$, $t_{ef}$, $r_{fe}$ and $t_{fe}$ represent reflection and transmission coefficients capturing the phenomenon of energy-exchange between wave components having different propagation characteristics. Such *wave-conversion* phenomenon is originated from the significantly different dispersive characteristics relative to distinct unit-cell regions (Figure S1). In particular, $r_{ef}$ and $t_{ef}$ map the amplitude change relative to the reflected/transmitted flexural wave components, originated from the evanescent ones that are incident towards Region A. Similarly, $r_{fe}$ and $t_{fe}$ capture the amplitude change relative to the reflected/transmitted evanescent wave components, originated from the propagating ones that are incident towards Region A. The described transmission and reflection coefficients can be found by applying suitable boundary conditions (*B.C*s) to Eq. (1). In particular, when assuming that the coupling of the rod with Region A can only occur through longitudinal vertical mechanical fields, the rod can only displace like a *piston* (pure *TE*). As a result, a uniform lateral displacement profile is expected in Region A. In such scenario, the *B.C*s shown in Eq. (*S11*) can be used to approximate the expected modal characteristics under the rod. In particular, the first three equations of Eq. (*S11*) map the equality in the displacement, slope and curvature at the edges of Region A.

$$t\left(-\frac{a}{2}\right) = w\left(\frac{a}{2}\right)$$
$$t'\left(-\frac{a}{2}\right) = w'\left(\frac{a}{2}\right)$$
$$t''\left(-\frac{a}{2}\right) = w''\left(\frac{a}{2}\right) \quad (S11)$$
$$w'''\left(-\frac{a}{2}\right) - t'''\left(\frac{a}{2}\right) = \frac{F}{E_t I_t} = \frac{Z_b vel(z=0)}{E_t I_t}$$

In contrast, the fourth equation captures the existence of a shear-force, at $S$, that counterbalances the laterally uniform force distribution at the bottom surface of the rod. After substituting Eqs. (2-3) (see the main manuscript) in Eq. (S11), the reflection and transmission coefficients, discussed above, can be found (see Eqs. (S12-S17)).

$$r = -\frac{(1-i)e^{-iak}F}{2F + (4+4i)E_t I_t k^3} \quad (S12)$$

$$t = \frac{\left(\frac{1}{2}+\frac{i}{2}\right)e^{-iak}(F + 4E_t I_t k^3)}{F + (2+2i)E_t I_t k^3} \quad (S13)$$

$$r_{ef} = t_{ef} = -\frac{(1-i)e^{(\frac{1}{2}-\frac{i}{2})ak}F}{2F + (4+4i)E_t I_t k^3} \quad (S14)$$

$$r_{fe} = t_{fe} = -\frac{(1+i)e^{(\frac{1}{2}-\frac{i}{2})ak}F}{2F + (4+4i)E_t I_t k^3} \quad (S15)$$

$$r_e = -\frac{(1+i)e^{ak}F}{2F + (4+4i)E_t I_t k^3} \quad (S16)$$

$$t_e = \frac{\left(\frac{1}{2}-\frac{i}{2}\right)e^{ak}(F + 4iE_t I_t k^3)}{F + (2+2i)E_t I_t k^3} \quad (S17)$$

It is worth pointing out that, when $F$ is equal to zero, thus indicating that the rod does not affect the propagation of flexural waves in the unit-cell, $[G]_{4x4}$ becomes (Eq. (S18)):

$$[G]_{4x4} = \begin{bmatrix} 0 & 0 & e^{-ika} & 0 \\ 0 & 0 & 0 & e^{ka} \\ e^{-ika} & 0 & 0 & 0 \\ 0 & e^{ka} & 0 & 0 \end{bmatrix} \quad (S18)$$

Therefore, as expected, in such simplified scenario, all the reflection coefficients, as well as the transmission coefficients associated to the process of wave-conversion (*i.e.* $t_{ef}$ and $t_{fe}$), become zero. Therefore, only the transmission coefficient (*t*) is not nulled and equal to its expected value after assuming Region A to only act as an acoustic delay line that phase shifts or attenuate any existing propagating and evanescent component of $t(x)$ and $w(x)$, by an amount that is proportional to $a$. It is now useful to manipulate $[G]_{4x4}$ in such a way that the amplitudes of the wave components coming from the left-side of the rod (*i.e.* $[t]_{4x1} = [t_l, t_{le}, t_r, t_{re}]^T$) become the independent variables of Eq. (S9), whereas those outgoing the right-side of the rod (*i.e.* $[w]_{4x1} = [w_l, w_{le}, w_r, w_{re}]^T$) act as the dependent ones. In such scenario, Eq. (S9) becomes:

$$\begin{bmatrix} w_l \\ w_{le} \\ w_r \\ w_{re} \end{bmatrix} = [C]_{4x4} \begin{bmatrix} t_l \\ t_{le} \\ t_r \\ t_{re} \end{bmatrix} \quad (S19)$$

In Eq. (S19), the matrix $[C]_{4x4}$, known as the *coupling matrix*, captures the wave transmission characteristics (from the left-side to the right-side of the rod). The derived expression for $[C]_{4x4}$ is reported in Eq. (S20).

$$[C]_{4x4} = \begin{bmatrix} -\dfrac{ie^{iak}(F+4ik^3)}{4k^3} & -\dfrac{ie^{\left(-\frac{1}{2}+\frac{i}{2}\right)ak}F}{4k^3} & -\dfrac{iF}{4k^3} & -\dfrac{ie^{\left(\frac{1}{2}+\frac{i}{2}\right)ak}F}{4k^3} \\ \dfrac{e^{\left(-\frac{1}{2}+\frac{i}{2}\right)ak}F}{4k^3} & \dfrac{e^{-ak}(F+4k^3)}{4k^3} & \dfrac{e^{\left(-\frac{1}{2}-\frac{i}{2}\right)ak}F}{4k^3} & \dfrac{F}{4k^3} \\ \dfrac{iF}{4k^3} & \dfrac{ie^{\left(-\frac{1}{2}-\frac{i}{2}\right)ak}F}{4k^3} & \dfrac{ie^{-iak}(F-4ik^3)}{4k^3} & \dfrac{ie^{\left(-\frac{1}{2}-\frac{i}{2}\right)ak}F}{4k^3} \\ -\dfrac{e^{\left(\frac{1}{2}+\frac{i}{2}\right)ak}F}{4k^3} & -\dfrac{F}{4k^3} & -\dfrac{e^{\left(\frac{1}{2}-\frac{i}{2}\right)ak}F}{4k^3} & -\dfrac{e^{ak}(F-4k^3)}{4k^3} \end{bmatrix} \quad (S20)$$

The derivation of $[C]_{4x4}$ is key to find the transmission matrix, $[T]_{4x4}$, relative to the entire unit-cell. However, in order to do so, it is necessary to apply an additional boundary condition that forces a periodic displacement distribution, with period equal to *L*, between adjacent periodic cells. This can be done by defining two new displacement vectors, $[t_-]_{4x1}$ and $[w_+]_{4x1}$, for *t*(x=-*L/2*) and *w*(x=*L/2*). In particular, when neglecting the existence of any loss mechanism, $[t_-]_{4x1}$ and $[w_+]_{4x1}$ are equal in magnitude and equivalent to modified versions of $[t]_{4x4}$ and $[w]_{4x4}$. Such modified versions are formed by phase-shifted or more attenuated copies of the propagating and evanescent components forming $[t]_{4x1}$ and $[w]_{4x1}$, respectively (see Eqs. (S21-S22)):

$$\begin{bmatrix} w_{l,+} \\ w_{le,+} \\ w_{r,+} \\ w_{re,+} \end{bmatrix} = [D]_{4x4} \begin{bmatrix} w_l \\ w_{le} \\ w_r \\ w_{re} \end{bmatrix} = \begin{bmatrix} e^{-i\phi} & 0 & 0 & 0 \\ 0 & e^{\phi} & 0 & 0 \\ 0 & 0 & e^{i\phi} & 0 \\ 0 & 0 & 0 & e^{-\phi} \end{bmatrix} \begin{bmatrix} w_l \\ w_{le} \\ w_r \\ w_{re} \end{bmatrix} \quad (S21)$$

$$\begin{bmatrix} t_{l,-} \\ t_{le,-} \\ t_{r,-} \\ t_{re,-} \end{bmatrix} = [D]_{4x4}^{-1} \begin{bmatrix} t_l \\ t_{le} \\ t_r \\ t_{re} \end{bmatrix} = \begin{bmatrix} e^{-i\phi} & 0 & 0 & 0 \\ 0 & e^{\phi} & 0 & 0 \\ 0 & 0 & e^{i\phi} & 0 \\ 0 & 0 & 0 & e^{-\phi} \end{bmatrix}^{-1} \begin{bmatrix} t_l \\ t_{le} \\ t_r \\ t_{re} \end{bmatrix} \quad (S22)$$

In Eqs. (S21-S22), $\phi$ is equivalent to $k \cdot (L+a)/2$ [1]. From Eqs. (S21-S22) it is possible to compute $[T]_{4x4}$, relative to all the wave components travelling between adjacent edges of the unit-cell. This can be done by using Eq. (S23).

$$[T]_{4x4} = [D]_{4x4} \cdot [C]_{4x4} \cdot [D]_{4x4} =$$

$$= \begin{bmatrix} \dfrac{e^{-ikL}(-iF + 4k^3)}{4k^3} & -\dfrac{ie^{(\frac{1}{2}-\frac{i}{2})kL}F}{4k^3} & -\dfrac{iF}{4k^3} & -\dfrac{ie^{(-\frac{1}{2}-\frac{i}{2})kL}F}{4k^3} \\ \dfrac{e^{(\frac{1}{2}-\frac{i}{2})kL}F}{4k^3} & \dfrac{1}{4}e^{kL}\left(4+\dfrac{F}{k^3}\right) & \dfrac{e^{(\frac{1}{2}+\frac{i}{2})kL}F}{4k^3} & \dfrac{F}{4k^3} \\ \dfrac{iF}{4k^3} & \dfrac{ie^{(\frac{1}{2}+\frac{i}{2})kL}F}{4k^3} & \dfrac{e^{ikL}(iF+4k^3)}{4k^3} & \dfrac{ie^{-ak-(\frac{1}{2}-\frac{i}{2})kL}F}{4k^3} \\ -\dfrac{e^{(-\frac{1}{2}-\frac{i}{2})kL}F}{4k^3} & -\dfrac{F}{4k^3} & -\dfrac{e^{(-\frac{1}{2}+\frac{i}{2})kL}F}{4k^3} & \dfrac{1}{4}e^{-kL}\left(4-\dfrac{F}{k^3}\right) \end{bmatrix} \quad (S23)$$

As a sanity check, it is useful to look at the value of $[T]_{4x4}$ when $F$ is set to be zero, thus when the rod does not perturb the propagation characteristics of the unit-cell. As evident, in such scenario, the expression of $[T]_{4x4}$ is heavily simplified (see Eq. (S24)), thus clearly mapping the case in which the unit-cell can only phase-shift or attenuate existing propagating and evanescent wave components, in the trench.

$$[T]_{4x4} = \begin{bmatrix} e^{-ikL} & 0 & 0 & 0 \\ 0 & e^{kL} & 0 & 0 \\ 0 & 0 & e^{ikL} & 0 \\ 0 & 0 & 0 & e^{-kL} \end{bmatrix} \quad (S24)$$

It is important to point out that, as expected, different frequency behaviors characterize corresponding components of $[T]_{4x4}$ and $[C]_{4x4}$. Such a unique feature, which is mostly determined by the nonlinear dependence of $k$ with respect to frequency (see Eq. (S6)), determines the existence of multiple not correlated frequencies ($f^{(n)}$) at which the rod exerts the largest influence on the propagation capability of the trench. In particular, from the analysis of the eigenvalues of $[T]_{4x4}$ (see Eq. (S23)), it is possible to determine the transmission coefficient, $T$.

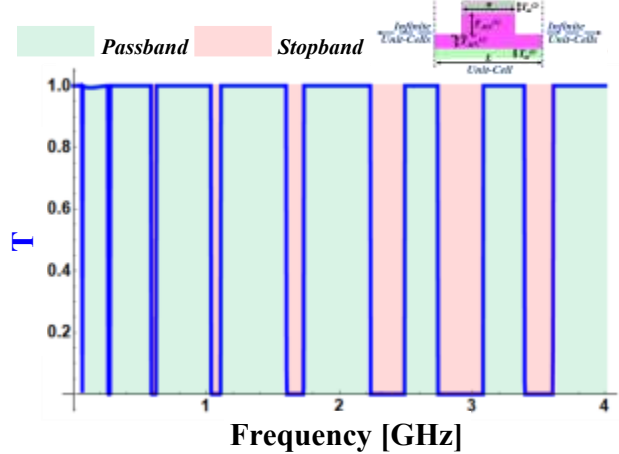

Figure S3: Simulated trend of $T$ vs. $f$, when assuming the same material and geometrical parameters used for considered unit-cell (see Figure S1) and when considering each edge connected to a in infinite sequence of unit cells.

As discussed in the main manuscript, this coefficient captures the reduction of the magnitude of a propagating displacement component, referring to one edge of an arbitrary chosen unit-cell and outgoing the same cell towards an infinite number ($N\to\infty$) of cascaded and identical unit-cells. As an example, $T$ is plotted, in Figure S3, when assuming the same unit-cell geometry and material stack adopted in our fabricated 2DRR. As evident, multiple stopbands exist for the propagation of the $A_0$ in the analyzed structure. It is straightforward to notice that the center frequencies of such *forbidden* bands closely match the $f^{(n)}$ (from 1 to 8) values, identified in Figure 2-a, in the main manuscript. This clearly proves that the largest influence of the rods, on the propagation characteristics of the unit-cell, occurs at those frequencies at which each rod produces virtual fixed-constraints across the lateral edges of the corresponding unit-cell.

In order to clearly visualize the evanescent behavior of the investigated unit-cell, an *ad-hoc* FEM simulation framework was created to analyze the propagation characteristics exhibited by a chain of unit-cells, for frequencies included in three expected stopbands. This framework uses different piezoelectric generators to produce longitudinal vibrations at significantly different frequencies (2.4GHz, 1.7GHz and 500MHz). This generator is attached, at one of its lateral side, to a perfectly-matched-layer (PML) while being connected, on the opposite side, to a chain of seven additional and electrically floating unit-cells (Figs. S4-S6). This chain acts as a delay line, separating the generator from an additional PML. We report, in **Error! Reference source not found.**-S6, the generated modeshape relative to the magnitude of the total-displacement across the chain of unit-cells, for the three investigated frequencies. Also, the distribution of the total displacement along a cut horizontal line, starting from the top-left edge of the first unit-cell of the chain and ending at the edge of the furthest PML, is also reported, for the same investigated cases.

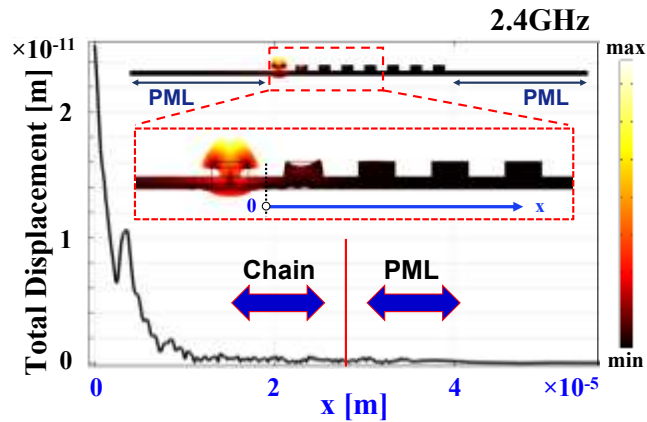

Figure S4: FEM simulated modeshape of the total displacement in a chain of 7 unit-cells (Fig. S1) when driven, at 2.4GHz, by one piezoelectric generator attached to it (here we used an electrically driven 2DRR unit-cell as the generator at 2.4GHz). The spatial total-displacement distribution along a cut line, starting from the top-left edge of the closest unit-cell to the generator and ending at the top-right corner of the furthest PML, is also report.

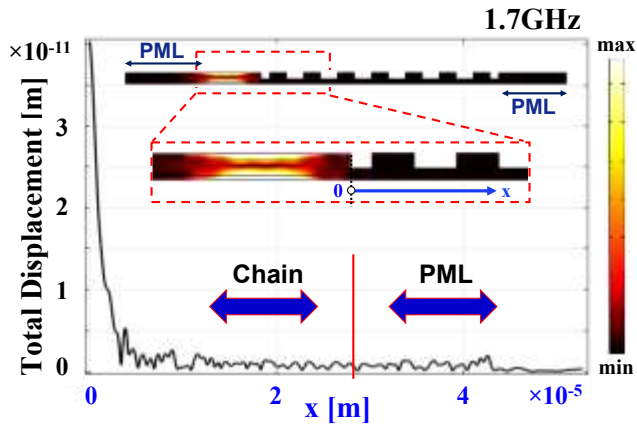

Figure S5: FEM simulated modeshape of the total displacement in a chain of 7 unit-cells (Fig. S1), when driven, at 1.7GHz, by one piezoelectric generator attached to it. The spatial total displacement distribution along a cut line, starting from the top-left edge of the closest unit-cell to the generator and ending at the top-right corner of the furthest PML, is also report.

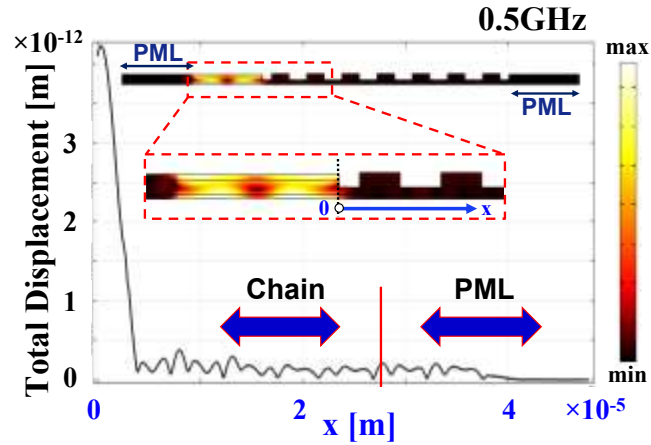

Figure S6: FEM simulated modeshape of the total displacement in a chain of 7 unit-cells (Fig. S1), when this is driven, at 0.5GHz, by one piezoelectric generator attached to it. The spatial total displacement distribution along a cut line, starting from the top-left edge of the closest unit-cell to the generator and ending at the top-right corner of the furthest PML, is also report.

As evident, the simulated displacement profiles, across the unit-cells and relative to the three investigated cases, clearly exhibit the typical exponential decay that is expected by a delay-line operating in its evanescent operational region.

## SII. Measured Impact of the Trenches on $k_t^2$

In order to experimentally demonstrate the impact of the trenches on the attainable $k_t^2$, a second device (Figure S7), with the same geometrical and material characteristics, but not relying on any trench, was simultaneously fabricated on the same silicon wafer than the reported 2DRR. As expected, this device showed a significantly lower $k_t^2$ (*<4.4%*) than attained by the fabricated 2DRR. Also, it showed a resistance at resonance ($R_{tot}=R_m+R_s$, being $R_m$ and $R_s$ its motional and series resistance, respectively), a loaded quality factor $Q_{3dB}$ (extracted from the 3dB bandwidth) and a $C_0$-value (being its static capacitance) in excess of *64Ω*, *236* and *330fF*, respectively. The measured $k_t^2$ and $C_0$ values for this device match closely their FEM predicted values (4.6%, 340*fF* respectively). Also, the measured $Q_{3dB}$-value for this device matches well the one attained by the reported 2DRR (see Figure 3 in the main manuscript). This fact clearly shows that the quality factor that we found for both this device and the 2DRR reported in the main manuscript is not limited by the presence of the trenches.

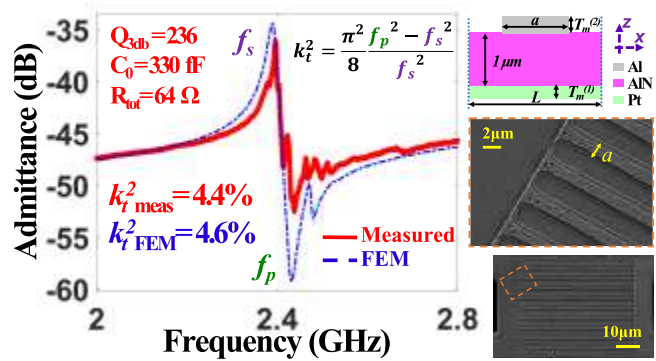

Figure S7: Measured (in red) and FEM simulated (in blue) admittance relative to a modified 2DRR device which was simultaneously fabricated on the same silicon wafer than the reported 2DRR and that uses the same geometrical and material characteristics of the fabricated 2DRR but not using any trench. A scanned-electron-microscope (SEM) picture of this fabricated device as well as a schematic representation of its cross-section are also shown.

## SIII. Fabrication Process

The fabrication of the presented 2DRR (Figure 1) follows the process flow shown in Figure S8. It starts with the sputtering deposition of a 250-nm-thick platinum full-plate, used as bottom electrode (Figure S8-a). It follows with the sputtering deposition of a 1-μm-thick AlN-film (Figure S8-b). Vias are then formed, in the AlN-layer, through a wet-etch process (Figure S8-c), followed by the sputtering and lift-off of a 330nm-thick aluminum layer, forming the top metal layer (Figure S8-d). Then, a 150-nm-thick gold layer is deposited, on the device pads and in the vias (Figure S8-e), so as to minimize their associated ohmic losses. Later, the releasing holes are formed through an AlN dry-etch (Figure S8-f), followed by the patterning of the top electrodes and the derivation of the AlN trenches (Figure S8-g). These two steps are attained through a combination of wet and dry-etch, which is optimized so as to minimize the surface roughness and optimize the AlN sidewall angle in the trenches. Finally, the devices are released through a XeF$_2$ silicon etching process (Figure S8-h).

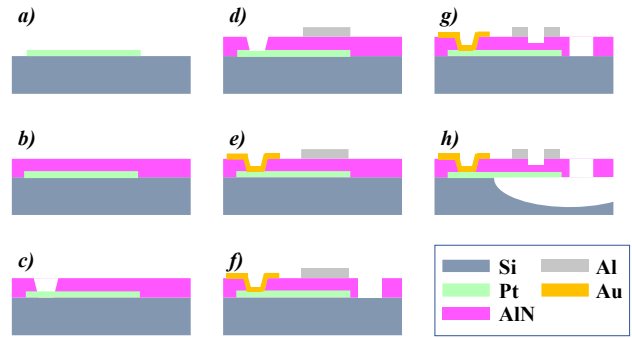

Figure S8: Fabrication process used for the reported 2DRR.